\documentclass[preprint,preprintnumbers,amsmath,amssymb,floatfix,nofootinbib,superscriptaddress,longbibliography]{revtex4-1}
\usepackage[T1]{fontenc} 
\usepackage{graphicx}
\usepackage{color}
\usepackage{hyperref}
\hypersetup{colorlinks=true,linkcolor=cyan,citecolor=cyan,filecolor=black,menucolor=black,urlcolor=cyan}
\usepackage{orcidlink}
\usepackage{easy-todo}
\usepackage{tikz}
\usetikzlibrary{decorations.markings}

\newcommand{\bk}{\boldsymbol{k}}
\newcommand{\bp}{\boldsymbol{p}}
\newcommand{\bq}{\boldsymbol{q}}

\begin{document}

\title{Classical Hamiltonian of Reissner-Nordstr\"om black holes at second post-Minkowskian order from scattering amplitudes}

\author{Allan Alonzo-Artiles\,\orcidlink{0009-0003-3202-8016}}
\email{allanea@estudiantes.fisica.unam.mx}
\author{Manfred Kraus\,\orcidlink{0000-0001-6668-6699}}
\email{mkraus@fisica.unam.mx}
\affiliation{Departamento de Física Teórica, Instituto de Física, \\
Universidad Nacional Autónoma de México, Cd. de México C.P. 04510, México}

\date{\today}

\begin{abstract}
We employ scattering amplitudes in Einstein-Maxwell theory to compute the
classical Hamiltonian of a binary system of two charged, non-spinning compact
objects. The effective Hamiltonian is valid to all orders in velocity and up to
second post-Minkowskian order (2PM), i.e. $\mathcal{O}(G^2)$.
The classical interaction potential is extracted via matching the four-point
tree and one-loop amplitudes to those of a suitably chosen classical effective
field theory.
We derive the 2PM scattering angle from the Hamiltonian and find complete
agreement with known results. Using also the third order result for the
scattering angle computed by Wilson-Gerow, we expand the scattering angle at
second post-Newtonian order (2PN). 
Additionally, we provide expressions for the binding energy and periastron
advance at 2PN order via the boundary-to-bound dictionary introduced by Kälin
and Porto and find agreement with previous findings.
\end{abstract}
\maketitle

\section{Introduction}
Understanding the relativistic two-body dynamics of compact binary mergers
remains a central challenge in gravitational wave physics.  Since the initial
direct detection of the first gravitational waves~\cite{LIGOScientific:2016aoc,
LIGOScientific:2016dsl, LIGOScientific:2016sjg, LIGOScientific:2017vwq,
LIGOScientific:2017bnn} by LIGO and Virgo and later joined by KAGRA the field
has transformed tremendously towards precision measurements, which allows us to
study the properties of black holes and perform consistency tests of general
relativity~\cite{LIGOScientific:2021sio}. Future upgrades of the
LIGO-Virgo-KAGRA detectors~\cite{KAGRA:2021vkt}, together with upcoming
observatories, such as Cosmic Explorer~\cite{Reitze:2019iox}, Einstein
Telescope~\cite{Punturo:2010zz}, LISA~\cite{LISA:2017pwj} and
TianQin~\cite{TianQin:2015yph}, will significantly enhance detector sensitivity
and thereby expand the potential to uncover hints of new
physics~\cite{Borhanian:2022czq}. To exploit the complete physics potential of
these observatories an improved theoretical modeling of gravitational waveforms
is necessary~\cite{Purrer:2019jcp}.

Even though most studies to date have focused on electrically neutral black
holes, there are compelling theoretical motivations to investigate compact
objects carrying a conserved charge. The presence of a charge has a profound
impact on the dynamics of the two-body system as in addition to the attractive
gravitational pull there can be another repulsive or attractive force at play.
This will modify the observed gravitational wave signals in all three stages,
the inspiral~\cite{Zhu:2018tzi}, the merger~\cite{Zhang:2016rli} and the
ringdown~\cite{Pani:2013ija, Pani:2013wsa, Zilhao:2014wqa, Mark:2014aja,
Dias:2015wqa, Dias:2021yju, Carullo:2021oxn}.

Our motivation to study the two-body dynamics of charged black holes is
twofold. First, within general relativity the No-hair
theorem~\cite{Israel:1967wq, Israel:1967za, Carter:1971zc, Hawking:1971vc}
guarantees that stationary black holes are characterized solely by three
fundamental properties: mass, spin and charge. However, in astrophysical
environments it is not expected that black holes can sustain any macroscopic
electrical charge for a long time~\cite{Gibbons:1975kk, Eardley:1975kp,
Blandford:1977ds}.  Nonetheless, even if the presence of electrically charged
black holes is not expected it would make future consistency tests of general
relativity more robust if the charge degree of freedom would be inferred from
data in the analysis of gravitational waves.

Secondly, many extensions of the Standard Model contain so-called dark sectors,
which modify dynamics through the presence of additional long-range
interactions that are associated with hidden gauge symmetries. These models
often give candidates for dark matter particles~\cite{DeRujula:1989fe,
Perl:1997nd, Cardoso:2016olt, Gupta:2021rod} or include dark photon
interactions~\cite{Feng:2008mu, Feng:2009mn}.  In these models, compact objects
may carry a dark U(1) charge and the same astrophysical restrictions that would
lead to the rapid neutralization of the electric charge do not apply in this
case. Furthermore, if compact objects beyond the conventional black holes and
neutron stars exist, such as boson stars, then those boson stars can carry
charge~\cite{Jetzer:1989av, Pugliese:2013gsa, Kain:2021bwd, Lopez:2023phk,
Mio:2025fnj} and mergers involving those objects therefore can showcase the
dynamics of a charged two-body system.

We will focus on the systematic improvement of the description of the inspiral
phase of charged black hole mergers. In comparison to Schwarzschild or Kerr
black holes their charged partners received only very little attention so far.
Most of the existing literature concerns the post-Newtonian (PN) expansion, in
which one simultaneously expands around weak gravitational fields, i.e. $Gm/(rc^2)\ll
1$, and small velocities, i.e $v^2/c^2\ll 1$, of the celestial objects. The first PN
corrections to the Lagrangian governing the dynamics of a charged binary black
hole system have been computed in Refs.~\cite{Julie:2017rpw, Khalil:2018aaj,
Patil:2020dme} as well as the dynamics in an external magnetic
field~\cite{Tang:2025jop}. The accuracy of the Lagrangian has been extended to
the second PN order in Refs.~\cite{Gupta:2022spq, Placidi:2025xyi}.
Furthermore, the radiation reaction has been recently computed at 2PN
accuracy~\cite{Verma:2025sdk}.  Focusing only on binary systems that exhibit
extreme mass ratios tidal effects have been studied in
Refs.~\cite{Pina:2022dye, Grilli:2024fds}.

An alternative approach is the so-called post-Minkowskian (PM) expansion in
which only weak fields are assumed, i.e $Gm/(rc^2) \ll 1$, but fully special
relativistic effects are taken into account. For instance, in
Ref.~\cite{Cheung:2020gbf} the dynamics of charged black holes have been
investigated in the probe limit to all orders in the PM expansion.
Furthermore, Ref.~\cite{Jones:2022aji} explored the Hamiltonian formulation of
many-body interactions. The article derives the three-body Hamiltonian at
second order in the PM expansion was derived.  We want to stress, that the
latter results depend on the two-body Hamiltonian, which however was not shown
explicitly in the article. In Ref. ~\cite{DOnofrio:2022cvn} the spherically
symmetric solutions for the metric and gauge field in Einstein-Maxwell theory
were derived from scattering amplitudes at 3PM order. Finally, the scattering
angle for charged binary black hole system was computed at the 3PM order in
Ref.~\cite{Wilson-Gerow:2023syq} using an effective field theory approach to
binary systems with extreme mass ratios.

The goal of this article is to present the calculation of the classical
Hamiltonian of charged binary black hole systems at second order in the PM
expansion using scattering amplitudes. We follow the methodology that has been
established in Refs.~\cite{FebresCordero:2022jts, Akpinar:2024meg,
Akpinar:2025bkt, Bohnenblust:2023qmy,  Bohnenblust:2025gir} by computing
directly the necessary gravitational scattering amplitudes in Einstein-Maxwell
theory. Our results show consistency with the 2PN results of
Ref.~\cite{Placidi:2025xyi}. It should be noted that to obtain full 2PN
accuracy we rely on the 3PM result for the scattering angle of Ref
.~\cite{Wilson-Gerow:2023syq}. Furthermore, for observables of bound-systems we
use the boundary-to-bound dictionary of
Refs.~\cite{Kalin:2019rwq,Kalin:2019inp}.

The article is organized as follows: In section~\ref{sec:einstein_maxwell} we
discuss the Einstein-Maxwell theory in the weak-field expansion as well as the
traditional black hole solution in unperturbed general relativity. In
section~\ref{sec:classical_limit} we discuss how classical terms can be
extracted from quantum scattering amplitudes via soft expansions. Afterwards,
in section~\ref{sec:eft} we elaborate how the classical two-body Hamiltonian
can be computed in terms of scattering amplitudes, followed by
section~\ref{sec:observables}, where we discuss classical observables and cross
checks. And finally, in section~\ref{sec:conclusions} we present our
conclusions and outlook.

\section{Einstein-Maxwell Field Theory}
\label{sec:einstein_maxwell}
In this section we will first discuss some properties of the induced spacetime
metric by the presence of a stationary charged black hole. Afterwards, we
develop the weak field expansion of the Einstein-Maxwell theory that we will
employ to compute perturbative scattering amplitudes in quantum field theory.

\subsection{The static Reissner-Nordstr\"om black hole solution}
The solution to the Einstein equations for a spherically symmetric, statically
charged, stationary black hole in Einstein-Maxwell
theory~\cite{Reissner:1916cle, Weyl:1917rtf, Nordstrom:2018acn,
10.1098/rspa.1921.0028} is given by
\begin{equation}
 ds^2 = \left(1-\frac{r_s}{r}+\frac{r^2_Q}{r^2}\right)dt^2 - \left(1-\frac{r_s}{r}+\frac{r^2_Q}{r^2}\right)^{-1}dr^2 - r^2 d\Omega^2\;,
 \label{eqn:Reissner-Nordstrom}
\end{equation}
where $r_s = 2GM$ is the Schwarzschild radius, $r^2_Q = Ge^2Q^2/(4\pi)$ is a
characteristic length scale associated with the electric charge $Q$ of the
black hole (in units of $e$), and $d\Omega^2$ is the volume element of the
two-sphere. An alternative representation of the Reissner-Nordström metric can
be obtained by factorizing the radial function as below
\begin{equation}
 1-\frac{r_s}{r}+\frac{r^2_Q}{r^2} = \left(1-\frac{r_+}{r}\right)\left(1-\frac{r_-}{r}\right)\;,
\end{equation}
where 
\begin{equation}
  r_{\pm} = \frac{1}{2}\left(r_s \pm \sqrt{r^2_s-4r^2_Q}\right) = GM \pm \sqrt{G^2M^2-\frac{Ge^2Q^2}{4\pi}}\;.
  \label{eqn:horizons}
\end{equation}
Consequently, Reissner-Nordstr\"om black holes exhibit two horizons if $2|r_Q|
< r_s$.  Here $r_+$ corresponds to the outer horizon and $r_-$ to the inner
one.  In the case where $2|r_Q| = r_s$ the outer and inner horizons coincide
and the electric charge of the black hole, called an extremal black hole,
reaches its maximum, which is given by
\begin{equation}
    eQ_{\textrm{max}} = \sqrt{4\pi G} M\;.
\end{equation}
For later convenience, we also present the form of the Reissner-Nordstr\"om
metric in isotropic coordinates. We use the following transformation given in
Ref.~\cite{Noble:2016cpq} to bring the metric in
Eq.~\eqref{eqn:Reissner-Nordstrom} to isotropic coordinates%
\begin{equation}
  r \mapsto r_{\textrm{iso}}\left(\left(1+\frac{r_s}{4r_{\textrm{iso}}}\right)^2-\frac{r^2_Q}{4r^2_{\textrm{iso}}}\right)\;.
  \label{eqn:isotropic map}
\end{equation}
The Reissner-Nordstr\"om metric then reads
\begin{equation}
 ds^2 = \left(\frac{\left(1-\frac{r_s}{4r}\right)\left(1+\frac{r_s}{4r}\right)+\frac{r^2_Q}{4r^2}}{\left(1+\frac{r_s}{4r}\right)^2-\frac{r^2_Q}{4r^2}}\right)^2dt^2 - \left(\left(1 + \frac{r_s}{4r}\right)^2 - \frac{r^2_Q}{4r^2}\right)^2(dr^2 + r^2d\Omega)
 \label{eqn:RN_isotropic}
\end{equation}
where we have dropped the subscript in $r_{\textrm{iso}}$ for notational
simplicity.
Finally, the electromagnetic four-potential $A_{\mu}(x)$ for a non-rotating
charged black hole is given by
\begin{equation}
  A_0(r) = \frac{eQ}{4\pi r}\;, \qquad A_i(r) = 0\;.
\end{equation}
Inserting Eq. \eqref{eqn:isotropic map} in the equation above we obtain the
expression for the four-potential in isotropic coordinates
\begin{equation}
  A_0(r) = \frac{eQ}{4\pi r \left(\left(1+\frac{r_s}{4r}\right)^2-\frac{r^2_Q}{4r^2}\right)}\;, \qquad A_i(r) = 0\;.
\label{eqn:four-potential}
\end{equation}
We will make use of the metric and four-potential in isotropic coordinates
later to perform a cross check of our results in the probe limit.
\subsection{Weak Field Expansion of Einstein-Maxwell theory}
We consider a two-body system of electrically charged, non-spinning black holes
with masses $m_1$ and $m_2$, and charges $eQ_1$ and $eQ_2$. The dynamics of the
system is governed by the Einstein-Maxwell Lagrangian minimally coupled to
matter:
\begin{equation}
    \mathcal{L} = \mathcal{L}_{\textrm{EH}} + \mathcal{L}_{\textrm{EM}} + \mathcal{L}_{m}\;,
\end{equation}
where $\mathcal{L}_{\textrm{EH}}$ is the Einstein-Hilbert Lagrangian,
$\mathcal{L}_\textrm{EM}$ is the electromagnetic Lagrangian, and
$\mathcal{L}_m$ represents the matter Lagrangian. Here we are considering the
point-particle approximation by neglecting finite-size
effects~\cite{Vaidya:2014kza}. This effective description is valid under the
assumption that the characteristic length scale $R$ of each black hole and the
separation $r$ between them satisfy the hierarchy of scales $R \ll r$.

The Einstein-Hilbert Lagrangian, describing pure gravitational interactions,
reads
\begin{equation}
    \mathcal{L}_\textrm{EH} = -\frac{2}{\kappa^2}\sqrt{-g}R\;,
\end{equation}
where $g = \det(g_{\mu\nu})$ is the determinant of the metric tensor
$g_{\mu\nu}$, $R$ is the Ricci scalar, and $\kappa = \sqrt{32\pi~G}$ is a
coupling related to Newton's constant $G$. We will consider an expansion of the
metric around a flat background space-time metric. The choice
\begin{equation}
  g_{\mu\nu} = \eta_{\mu\nu} + \kappa h_{\mu\nu}\;,
  \label{eqn:weak-field}
\end{equation}
defines the \textit{weak field expansion} and $h_{\mu\nu}$ is identified with
quantum fluctuations in the gravitational field, i.e. the dynamic graviton
quantum field. The expansion generates infinite many terms through
\begin{equation}
\begin{split}
 g^{\mu\nu} &= \eta^{\mu\nu} - \kappa h^{\mu\nu} + \kappa^2 h^{\mu\lambda}h^\nu_{~\lambda}  + \cdots\;, \\
 \sqrt{-g} &= 1 + \frac{\kappa}{2}h + \frac{\kappa^2}{8}\Big(h^2 - 2h_{\mu\nu}h^{\mu\nu}\Big) + \cdots\;,
\label{eqn:infinite-expansion}
\end{split}
\end{equation}
where we have defined $h\equiv h^\lambda_{~\lambda}$. The quantization
procedure of the graviton field $h_{\mu\nu}$ requires to supplement the
Lagrangian with a gauge-fixing term, where we have chosen the commonly used
\textit{linearized harmonic gauge}. The corresponding Lagrangian is explicitly
given by
\begin{equation}
 \mathcal{L}_\textrm{GF-GR} = \eta^{\mu\nu} \left(\partial^\lambda h_{\mu\lambda} 
- \frac{1}{2}\partial_\mu h\right) \left(\partial^\lambda 
 h_{\nu\lambda} - \frac{1}{2}\partial_\nu h\right)\;,
\end{equation}
where indices have been contracted using the flat space-time metric
$\eta_{\mu\nu}$. In this gauge the $D$-dimensional graviton propagator takes
the form~\cite{Bern:2002kj}
\begin{equation}
 P_{\mu\nu,\alpha\beta} = \frac{i}{p^2+i\epsilon}~\frac{1}{2}\left[
 \eta_{\mu\alpha}\eta_{\nu\beta} + \eta_{\mu\beta}\eta_{\nu\alpha} -
 \frac{2}{D-2}\eta_{\mu\nu}\eta_{\alpha\beta}\right]\;.
 \label{eqn:grav_prop}
\end{equation}
The electromagnetic Lagrangian is given by
\begin{equation}
\frac{\mathcal{L}_{\textrm{EM}}}{\sqrt{-g}} = -\frac{1}{4}g^{\mu\alpha}g^{\nu\beta}F_{\mu\nu}F_{\alpha\beta} - \frac{1}{2\xi}(g^{\mu\nu}\nabla_\mu A_\nu)^2\;,
\end{equation}
where $\xi$ is the gauge-fixing parameter and $\nabla_\mu A_\nu = \partial_\mu
A_\nu - \Gamma^\lambda_{~\mu\nu}A_\lambda$ is the covariant derivative of
general relativity.  Furthermore, the electromagnetic field strength tensor is
given by $F_{\mu\nu} = \nabla_\mu A_\nu - \nabla_\nu A_\mu = \partial_\mu A_\nu
- \partial_\nu A_\mu$. Lastly, the matter Lagrangian includes massive, complex
scalar fields $\phi_i$ which couple minimally to $g_{\mu\nu}$ and to $A_{\mu}$
via their electric charge 
\begin{equation}
 \frac{\mathcal{L}_{m}}{\sqrt{-g}}= \sum_{i=1}^2\left[ g^{\mu\nu}(D_\mu\phi_i)^\dagger(D_\nu\phi_i) - m_i^2\phi_i^\dagger\phi_i\right]\;,
\end{equation}
where $D_\mu = \partial_\mu - ieQ_iA_\mu$ is the covariant derivative of
quantum electrodynamics.

\section{Scattering Amplitudes and their Classical Limit}
\label{sec:classical_limit}
The classical dynamics of a binary system of non-rotating charged black holes
are encapsulated in the scattering process of two massive complex scalar fields
$\phi_{1,2}$ with masses $m_{1,2}$ and charges $eQ_{1,2}$, respectively
\begin{equation}
    \phi_1(-p_1) + \phi_2(-p_2) \to \phi_1(p_4) + \phi_2(p_3)\;,
\end{equation}
where the signs for the external momenta follow an all-outgoing convention. 
Following Ref.~\cite{Parra-Martinez:2020dzs} we parametrize the momenta by
special kinematic variables that facilitate the extraction of classical
physics. The explicit parametrization is given by
\begin{equation}
\begin{gathered}
    p_1 = -\bar{p}_1 + q/2\,, \quad p_4 = \bar{p}_1 + q/2\;, \\
    p_2 = -\bar{p}_2 - q/2\,, \quad p_3 = \bar{p}_2 - q/2\;,
\end{gathered}
\label{barred momenta}
\end{equation}
for which $\bar{p}_i \cdot q = 0$ by construction, and where $q = p_1 + p_4$ is
the momentum transfer. We further define normalized velocity vectors $u_i$
through the relations $\bar{p}_i = \bar{m}_iu_i$ and $\bar{m}_i^2 = m^2_i
-q^2/4$, and the dimensionless parameter $y = u_1 \cdot u_2$, which is a
measure of the relative velocity between the two scattered particles. 

In accordance with the correspondence principle, the classical limit can be
defined as the limit of large conserved charges. In the context of black hole
scattering, we have large orbital angular momentum, masses and center-of-mass
energy. This implies the hierarchy of scales~\cite{Parra-Martinez:2020dzs}
\begin{equation}
  m^2_{\star} \sim s, |u|, m_1^2, m^2_2 \sim J^2 |t| \gg |t| = |q|^2\;,
\end{equation}
where $m_{\star}$ is some characteristic mass scale and $s$, $t$, and $u$
Mandelstam variables. Thus, the classical dynamics are defined by an expansion
for small $|q|$. From the hierarchy above, the phase space can be divided into
two momentum regions defined based on the possible scalings of the momentum
components $k = (\omega, \bk)$ of a given (external or loop) momentum
\begin{equation}
\begin{split}
    \textrm{hard:}\ &(\omega, \bk) \sim (m_{\star},m_{\star})\,, \\
    \textrm{soft:}\ &(\omega, \bk) \sim (|q|,|q|)\;.
\end{split}
\end{equation}
Following Ref.~\cite{Bern:2019crd}, we can further subdivide the soft region
through an additional small parameter $v \ll 1$, related to $y \approx
1/\sqrt{1-v^2}$. We then have the following subregions inside the soft region
\begin{equation}
\begin{split}
    \textrm{potential:}\ &(\omega, \bk) \sim (|q|v,|q|)\;, \\
    \textrm{radiation:}\ &(\omega, \bk) \sim (|q|v,|q|v)\;.
\end{split}
\end{equation}
For our purposes the classical conservative dynamics arise from interactions
mediated by photons and gravitons in the potential region.  The reason is
because potential modes are off-shell and hence can be integrated out and its
effects captured through an instantaneous contact-interaction potential in an
effective field theory description. It is known that the strict separation
of potential and radiation region breaks down at higher post-Minkowskian
orders~\cite{Bern:2021yeh}.

On dimensional grounds, let us consider the relevant length scales in the
scattering process 
\begin{equation}
\begin{split}
   &\textrm{Compton wavelength:}\ \ell_c \sim \frac{\hbar}{m}\;, \\
   &\textrm{Outer horizon radius:}\ r_+ \sim Gm + \sqrt{G^2m^2-Ge^2Q^2}\;, \\
   &\textrm{Impact parameter:}\ |b| \sim \frac{\hbar}{|q|}\;, 
\end{split}
\end{equation}
where $m \sim m_1 + m_2$ is some common mass scale, and the impact parameter
$|b|$ is the Fourier conjugate to the momentum transfer $|q|$. The distance
$r_+$ is bounded as $r_s/2 \leq r_+ \leq r_s$, the lower bound given by the
extremal limit $eQ \to \sqrt{4\pi G}m$ and the upper bound given by the limit
of vanishing charge $Q \to 0$. Hence, we can make the replacement $r_+ \to r_s$
when considering the characteristic size of the charged black holes without any
loss of generality. With this in mind, the PM expansion is established from the
following hierarchy of scales~\cite{DiVecchia:2021bdo}
\begin{equation}
  \ell_c\, \ll\, r_s\, \ll\, |b| \quad \Longleftrightarrow \quad 
  \frac{\hbar}{m}\, \ll\, Gm \,\ll\, \frac{\hbar}{|q|}\;,
  \label{eqn:length_scales}
\end{equation}
which implies the dimensionless ratios
\begin{equation}
    \frac{\ell_c}{|b|} \ll \frac{r_s}{|b|} \ll 1 \quad \Longleftrightarrow \quad \frac{|q|}{m} 
    \ll \frac{Gm|q|}{\hbar} \ll 1 \;.
\end{equation}
The PM expansion is defined as an expansion in $r_s/|b| \sim Gm/|b| \sim Gm|q|$
(in natural units). Therefore, at each order in $G$, we need to expand the
amplitude to an additional order in $|q|$ in the small-$|q|$ expansion in order
to extract the classical dynamics. Specifically, at $\mathcal{O}(G^n)$ the
scattering amplitude only needs to be expanded up to $\mathcal{O}(|q|^{n-3})$.

There is an additional length scale that we can construct, which is given by
the classical charge radius $r_{\textrm{charge}} = e^2Q^2/(4\pi m)$. Multiplying
this scale by one through the ratio $G/G$ we can establish the relation
\begin{equation}
  r_{\textrm{charge}} = \frac{Ge^2Q^2}{4\pi} \frac{1}{Gm}\,\leq\, G^2m^2 \frac{1}{Gm} \sim r_s\;,
\end{equation}
where the inequality above follows from the relation $2|r_Q|\leq r_s$ required
for physically relevant charged black hole solutions (see Eq.
\eqref{eqn:horizons}). If we define the dimensionless charge-to-mass ratio
$\eta = eQ/(\sqrt{4\pi G}m)$, then we see that $r_{\textrm{charge}} = Gm\eta^2
= r_s\,\eta^2$. The value of $|\eta|$ ranges from $0$ in the vanishing charge
limit to $1$ in the extremal limit. It follows that
\begin{equation}
  \frac{r_{\textrm{charge}}}{|b|}\, \leq\, \frac{r_s}{|b|}\, \ll\, 1 \quad \Longleftrightarrow 
  \quad \frac{Gm\eta^2|q|}{\hbar}\, \leq\, \frac{Gm|q|}{\hbar}\, \ll\, 1\;.
\end{equation}
An expansion in orders of the small ratio $r_{\textrm{charge}}/|b| \ll 1$
defines the so-called post-Lorentzian (PL) regime \cite{Bern:2021xze,
Bern:2023ccb}. The above implies that the classical PL regime is automatically
captured when enforcing the hierarchy of scales in
Eq.~\eqref{eqn:length_scales} of the classical PM regime. Therefore, we only
need to concern ourselves with the PM expansion in orders of the small ratio
$r_s/|b| \ll 1$.

\begin{figure}[t]
 \centering
 \begin{minipage}{0.24\textwidth}
  \includegraphics[width=\textwidth]{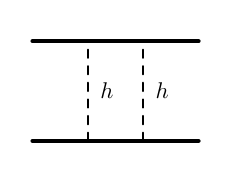}\\ (a) -- $G^2$
 \end{minipage}
 \begin{minipage}{0.24\textwidth}
  \includegraphics[width=\textwidth]{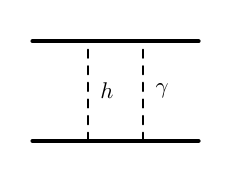}\\ (b) -- $G^2\eta_1\eta_2$
 \end{minipage}
 \begin{minipage}{0.24\textwidth}
  \raisebox{0.6cm}{\includegraphics[width=\textwidth]{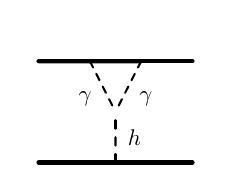}}\\ (c) -- $G^2\eta_i^2$
 \end{minipage}
 \begin{minipage}{0.24\textwidth}
  \includegraphics[width=\textwidth]{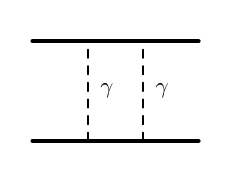}\\ (d) -- $G^2\eta_1^2\eta_2^2$
 \end{minipage}
 \caption{Representative diagrams for the various gauge-invariant building
block of the one-loop scattering amplitude in Einstein-Maxwell theory.}
 \label{fig:amps}
\end{figure}

The necessary scattering amplitude can be decomposed into four classes of
gauge-invariant building blocks parameterized by powers of $\eta_i$, as shown
in Figure~\ref{fig:amps}. For charged black holes the scattering amplitude at
$\mathcal{O}(G^2)$, i.e at 2PM, involves pure gravitational and pure
electromagnetic interactions as shown in Fig.\ref{fig:amps} (a) and (d).  Note
that, due to the definition of $\eta_i = eQ_i/(\sqrt{4\pi G}m_i)$ the pure
electromagnetic contribution in (d) does not depend on Newton's coupling, since
$G^2\eta_1^2\eta_2^2 \sim G^2e^2Q_1^2Q_2/G^2 = e^2Q_1^2Q_2^2$.  Furthermore, at
this order we find, as depicted in Fig.~\ref{fig:amps} (b), the first mixed
contributions with graviton and photon exchanges. Finally, in
Fig.~\ref{fig:amps} (c) we find corrections that do not have an underlying
tree-level contribution and are due to the gravitational field interacting
directly with the energy density of the electromagnetic field.

To compute the scattering amplitudes we obtain the necessary Feynman rules for
Einstein-Maxwell theory using the weak field expansion of the Lagrangian of
section~\ref{sec:einstein_maxwell} using the \textsc{xAct}
packages~\cite{Brizuela:2008ra,Nutma:2013zea}. We implemented the obtained
rules in \textsc{Caravel}~\cite{Abreu:2020xvt}, which implements a numerical
variant of the generalized unitarity method~\cite{Ita:2015tya, Abreu:2017idw,
Abreu:2017xsl, Abreu:2017hqn, Abreu:2018jgq}.  We evaluated the one-loop
scattering amplitudes numerically using finite modular
arithmetics~\cite{vonManteuffel:2014ixa, Peraro:2016wsq}, which allows the
reconstruction of the complete analytical expression of the quantum scattering
amplitudes. Afterwards, we expand the amplitudes in the potential region to
extract the classical contributions.

The final expressions for the 1PM and 2PM classical scattering amplitudes in
the center-of-mass frame are given by:
\begin{align}
    \mathcal{M}_1 &= -\frac{4\pi G \nu^2 m^2}{\gamma^2 \xi \bq^2}(1-2\sigma^2+\eta_1\eta_2\sigma) \;, \\
    \mathcal{M}_2 &= \frac{\pi^2 G^2 \nu^2 m^3}{2\gamma^2 \xi |\bq|}\Bigg[-3(1-5\sigma^2) + 2\eta^2_1\eta^2_2-12\eta_1\eta_2\sigma+\Big(\frac{\eta^2_1+\eta^2_2}{2}+\frac{\eta^2_1-\eta^2_2}{2}\sqrt{1-4\nu}\Big)(1-3\sigma^2)\Bigg] \\ \nonumber
    &\quad +\frac{1}{\sqrt{\sigma^2-1}}\frac{2\pi i \nu^3 m^4 G^2}{\gamma^2 \xi \bq^2}(1-2\sigma^2+\eta_1\eta_2\sigma)^2g(\epsilon) \;.
    \label{eqn:2PM_amplitude}
\end{align}
In writing the above expressions, we have included the non-relativistic
normalization factor $1/(4E_1E_2)$, where $E_{i} = \sqrt{\bp^2+m^2_{i}}$. For
convenience we also introduced the total energy $E$, the symmetric energy ratio
$\xi$, the total mass $m$, the symmetric mass ratio $\nu$, the energy-mass
ratio $\gamma$, and the relativistic invariant $\sigma$ as
\begin{align}
    E &= E_1+E_2\;, \quad \xi = \frac{E_1E_2}{(E_1+E_2)^2}\;, 
    \quad \sigma = \frac{p_1 \cdot p_2}{m_1 m_2}\;,\\
    m &= m_1+m_2\;, \quad \nu = \frac{m_1m_2}{(m_1+m_2)^2}\;, \quad \gamma = \frac{E}{m}\;.
\end{align}
Without loss of generality, we choose $m_1 > m_2$ such that
\begin{equation}
 m_1 = \frac{m}{2}\left[1 + \sqrt{1-4\nu}\right]\;, \qquad
 m_2 = \frac{m}{2}\left[1 - \sqrt{1-4\nu}\right]\;.
\end{equation}
We also define 
\begin{equation}
    g(\epsilon) \equiv \left(\frac{-q^2}{\bar{\mu}^2}\right)^{-\epsilon}e^{\gamma_E \epsilon}\frac{\Gamma(-\epsilon)^2\Gamma(1+\epsilon)}{\Gamma(-2\epsilon)} \;,
\end{equation}
where $\bar{\mu}^2 = 4\pi e^{-\gamma_E}\mu^2$ and $\mu$ is the dimensional
regularization scale. The term in $\mathcal{M}_2$ proportional to $g(\epsilon)$
is divergent which reflects the IR divergence of the original integral that
originated the term. It is also proportional to
$(1-2\sigma^2+\eta_1\eta_2\sigma)^2$ and represents the double iteration of the
tree-level diagram. In the next section, we will see that this IR divergence
will cancel with a similar contribution originating from the effective theory,
that we employ to extract the classical interaction potential.

\section{Effective Field Theory}
\label{sec:eft}
Our goal is to extract from the 1PM and 2PM scattering amplitudes the classical
Hamiltonian for interacting spinless particles. As mentioned earlier, the
amplitudes contain infrared divergent and superclassical contributions that
need to be explicitly cancelled. To do so we construct an effective field theory
(EFT) that includes only the relevant dynamical degrees of freedom and captures
the classical physics of the considered $2 \to 2$ scattering process.  Then the
Hamiltonian can simply be extracted by matching the EFT to the full theory
results.

The physics we want to capture is the \textit{elastic} scattering of
\textit{positive} energy states. Since we are interested in conservative
dynamics and want to describe interactions via some classical potential, we
also impose that interactions be instantaneous. The EFT that describes the
physics of two non-relativistic (i.e. no anti-particles) massive scalar fields
interacting through a long range potential was introduced in Ref.
\cite{Cheung:2018wkq}. In the center-of-mass frame, the EFT Lagrangian in
momentum space is given by
\begin{multline}
  \mathcal{L} = \sum_{i=1}^{2} \int \frac{\mathrm{d}^{D-1}\boldsymbol{k}}{(2\pi)^{D-1}}\phi^{\dagger}_i(-\boldsymbol{k})
  \left(i\partial_t - \sqrt{\boldsymbol{k}^2+m^2_i}\right)\phi_i(\boldsymbol{k}) \\
  - \int \frac{\mathrm{d}^{D-1}\boldsymbol{k}}{(2\pi)^{D-1}}\frac{\mathrm{d}^{D-1}\boldsymbol{k'}}{(2\pi)^{D-1}} V(\boldsymbol{k},\boldsymbol{k'})
  \phi^{\dagger}_1(\boldsymbol{k'})\phi_1(\boldsymbol{k})\phi^{\dagger}_2(-\boldsymbol{k'})\phi_2(-\boldsymbol{k})\;.
\label{eqn:EFT}
\end{multline}
This effective field theory can be obtained from the Einstein-Maxwell theory
minimally coupled to matter by integrating out the matter negative energy
states and the potential modes of photons and gravitons. As a result, the
potential, which appears as some contact interaction, is non-local in space and
can be thought of as a Wilson coefficient. Following Ref. \cite{Bern:2021xze},
the form of the potential in momentum space is written in a PM expansion as
\begin{equation}
\begin{split}
  V(\boldsymbol{k},\boldsymbol{k'}) &= \sum_{n=1}^{\infty} \frac{(G/2)^n (4\pi)^{(D-1)/2}}{|\boldsymbol{k}-\boldsymbol{k'}|^{D-1-n}}
  \frac{\Gamma[(D-1-n)/2]}{\Gamma(n/2)}c_n\Big(\frac{\boldsymbol{k}^2+\boldsymbol{k'}^2}{2}\Big) \\
  &= \frac{4\pi G}{|\boldsymbol{k}-\boldsymbol{k'}|^2}c_1\Big(\frac{\boldsymbol{k}^2+\boldsymbol{k'}^2}{2}\Big)  
  + \frac{2\pi^2 G^2}{|\boldsymbol{k}-\boldsymbol{k'}|}c_2\Big(\frac{\boldsymbol{k}^2+\boldsymbol{k'}^2}{2}\Big) + \cdots \;,
\end{split}
\label{eqn:potential}
\end{equation}
where the $c_n$ are free coefficients that are fixed by matching to the full
theory amplitudes and encode the conservative dynamics. The various
normalization factors in Eq.~\eqref{eqn:potential} are chosen such that the
expression of the potential in position-space is simplifed. The resulting
Feynman rules for the non-relativistic propagator and interaction vertex can be
obtained from Eq. \eqref{eqn:EFT} and are given by
\begin{equation}
\raisebox{-0.5cm}{\includegraphics[height=1.75cm]{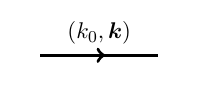}} =  \frac{i}{k_0-\sqrt{\boldsymbol{k}^2 + m^2_i}+i \varepsilon} \;,  \qquad
\raisebox{-1.2cm}{\includegraphics[height=2.5cm]{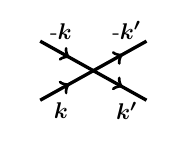}}  = -i V(\boldsymbol{k},\boldsymbol{k'}) \;.
\label{eqn:EFTrules}
\end{equation}

\begin{figure}[t]
\includegraphics[height=3.5cm]{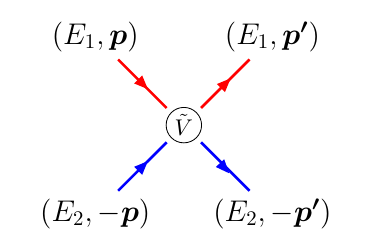}
\raisebox{1.5cm}{\Large \textbf{+}}
\includegraphics[height=3.5cm]{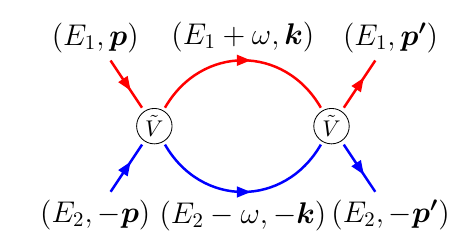}
\caption{The tree-level and one-loop EFT diagram necessary for the matching.
The loop momentum is given by $\ell^\mu = (\omega,\boldsymbol{\ell})$ and
therefore $\boldsymbol{k} = \boldsymbol{p}+\boldsymbol{\ell}$.}
\label{fig:EFT_diags}
\end{figure}

The amplitude in the effective field theory consists out of iterated bubble
diagrams and up to the one-loop order all relevant diagrams are shown in
Fig.~\ref{fig:EFT_diags}. Here, we assume $|\boldsymbol{p| =
|\boldsymbol{p'}|}$ for elastic scattering and energies are given by $E_i =
\sqrt{\boldsymbol{p}^2 + m^2_i}$. In the center-of-mass frame, the momentum
transfer $q = p_1 + p_4$ is purely spatial and is given by $\boldsymbol{q} =
\boldsymbol{p-\boldsymbol{p'}}$. 

To perform the EFT amplitude computations it is convenient to integrate the two
non-relativistic propagators in a bubble over the energy component of the loop
momentum. This yields an effective ``two-body'' propagator
\begin{equation}
\begin{split}
    i\Delta(\boldsymbol{k}) &\equiv \int \frac{\mathrm{d}\omega}{2\pi} 
    \frac{i}{E_1+\omega-\sqrt{\boldsymbol{k}^2+m^2_1}} 
    \frac{i}{E_2-\omega-\sqrt{\boldsymbol{k}^2+m^2_2}} \\
&=  \frac{i}{E_1+E_2-\sqrt{\boldsymbol{k}^2+m^2_1}-\sqrt{\boldsymbol{k}^2+m^2_2}}\;,
\end{split}
\end{equation}
where the Feynman $i\varepsilon$-prescription is left implicit.  The 1PM
amplitude in the effective theory is given by the tree-level diagram, which
involves only the $\mathcal{O}(G)$ term of the potential in Eq.
\eqref{eqn:potential}, and therefore
\begin{equation}
    \mathcal{M}^{\textrm{EFT}}_1 = -V(-\boldsymbol{p},-\boldsymbol{p'}) = -\frac{4\pi G}{\boldsymbol{q}^2}c_1(\boldsymbol{p}^2) \;,
\end{equation}
where we have set $\boldsymbol{k}-\boldsymbol{k'} =
-\boldsymbol{p}+\boldsymbol{p'} = -\boldsymbol{q}$ and $\boldsymbol{k}^2 =
\boldsymbol{k'}^2 = \boldsymbol{p}^2$ in
$c_1(\frac{\boldsymbol{k}^2+\boldsymbol{k'}^2}{2})$ for on-shell kinematics. In
the above, the first argument of the potential corresponds to the three-momenta
of $p_1$, and we had to change from an outgoing to an incoming convention given
how the Feynman interaction vertex is defined, and the second argument
corresponds to the three-momenta of $p_4$. 

The 2PM amplitude receives contributions from two diagrams: the tree-level
diagram given by the contact interaction vertex, keeping only the
$\mathcal{O}(G^2)$ term of the potential in Eq. \eqref{eqn:potential} and the
one-loop bubble diagram with only the $\mathcal{O}(G)$ term of the potential
inserted at each vertex
\begin{equation}
\begin{split}
    \mathcal{M}^{\textrm{EFT}}_2 &= -\frac{2\pi^2G^2}{|\boldsymbol{q}|}c_2(\boldsymbol{p}^2) - \int \frac{\mathrm{d}^{D-1}\boldsymbol{\ell}}{(2\pi)^{D-1}} V(-\boldsymbol{p},\boldsymbol{\ell})
    \Delta(\boldsymbol{\ell}) V(\boldsymbol{\ell},-\boldsymbol{p'}) \\
    &= -\frac{2\pi^2G^2}{|\boldsymbol{q}|}c_2(\boldsymbol{p}^2) - \int 
    \frac{\mathrm{d}^{D-1}\boldsymbol{\ell}}{(2\pi)^{D-1}} \frac{4\pi G}{|-\boldsymbol{p}-\boldsymbol{\ell}|^2}c_1\left(\frac{\boldsymbol{p}^2+\boldsymbol{\ell}^2}{2}\right) 
    \Delta(\boldsymbol{\ell})\frac{4\pi G}{|\boldsymbol{\ell}+\boldsymbol{p'}|^2}c_1\left(\frac{\boldsymbol{\ell}^2+\boldsymbol{p'}^2}{2}\right) \;.
\end{split}
\end{equation}
By changing integration variables $\boldsymbol{\ell} \to
\boldsymbol{\ell}-\boldsymbol{p}$, we arrive at the expression
\begin{equation}
    \mathcal{M}^{\textrm{EFT}}_2 =
    -\frac{2\pi^2G^2}{|\boldsymbol{q}|}c_2(\boldsymbol{p}^2) - 16\pi^2G^2\int 
    \frac{\mathrm{d}^{D-1}\boldsymbol{\ell}}{(2\pi)^{D-1}} \frac{1}{\boldsymbol{\ell}^2 (\boldsymbol{\ell}-\boldsymbol{q})^2}c^2_1\left(\frac{\boldsymbol{p}^2+(\boldsymbol{\ell}-\boldsymbol{p})^2}{2}\right)
    \Delta(\boldsymbol{\ell}-\boldsymbol{p}) \;.
\label{eqn:2PM_EFT}
\end{equation}
For off-shell momenta, the potential $V(\boldsymbol{k},\boldsymbol{k'})$ in the
loop integral will introduce quantum corrections which, of course, do not
contribute to the classical dynamics. Therefore, we have to expand the
integrand in the classical limit. For this we first define the variable
\begin{equation}
    Y \equiv \boldsymbol{\ell}^2 -2\boldsymbol{\ell}\cdot\boldsymbol{p}\;,
\end{equation}
so that we obtain
\begin{equation}
c^2_1\left(\frac{\boldsymbol{p}^2+(\boldsymbol{\ell}-\boldsymbol{p})^2}{2}\right)
= c^2_1\left(\boldsymbol{p}^2+\frac{Y}{2}\right)\;,
\end{equation}
and the two-body propagator can be written as
\begin{equation}
    \Delta(\boldsymbol{\ell}-\boldsymbol{p}) = \frac{1}{E-\sqrt{E^2_1+Y}-\sqrt{E^2_2-Y}}\;.
\end{equation}
In the classical limit $|\ell| \sim |q|$ and we can see that $Y$ does not have
a homogeneous $|q|$-scaling. Nevertheless, the leading-order piece in $Y$ is of
order $|q|$ so in the classical limit enforced by the hierarchy of scales
$|\ell| \sim |q| \ll |p_i|, m_i$ we can expand both the two-body propagator and
the coefficient function $c_1$ around $Y$. We obtain the following expansions
\begin{equation}
\begin{split}
    \Delta(\boldsymbol{\ell}-\boldsymbol{p}) &= -\frac{2\xi E}{Y} - \frac{(1-3\xi)}{2\xi E} + \ldots\;,\\
    c^2_1\left(\boldsymbol{p}^2+\frac{Y}{2}\right) &= c^2_1(\boldsymbol{p}^2) + Y c_1(\boldsymbol{p}^2)c'_1(\boldsymbol{p}^2) + \ldots \;,
\end{split}
\end{equation}
where $c'_1$ denotes the first derivative of $c_1$ with respect to
$\boldsymbol{p}^2$. We plug the above expansion back in Eq. \eqref{eqn:2PM_EFT}
and keep the terms which scale up to $\mathcal{O}(q)$, which is consistent with
the potential in Eq. \eqref{eqn:potential} up to $\mathcal{O}(G^2)$. The
resulting $(D-1)$-dimensional integrals have been computed in dimensional
regularization and are in full agreement with those reported in
Ref.~\cite{Parra-Martinez:2020dzs}. The final expressions for the EFT
amplitudes up to 2PM order are given by
\begin{align}
    \mathcal{M}^{\textrm{EFT}}_1 &= -\frac{4\pi G c_1}{\boldsymbol{q}^2}\;, \\
    \mathcal{M}^{\textrm{EFT}}_2 &= -\frac{2\pi^2G^2c_2}{|\boldsymbol{q}|} + \frac{\pi^2 G^2}{E\xi |\boldsymbol{q}|}\Bigg[(1-3\xi)c^2_1+4\xi^2E^2c_1c'_1\Bigg] + \frac{1}{\sqrt{\sigma^2-1}}\frac{2\pi i \xi E^2G^2c^2_1}{m^2 \nu\boldsymbol{q}^2}g(\epsilon) \;.
    \label{eqn:EFT_2PM_amplitude}
\end{align}
Since the EFT in Eq. \eqref{eqn:EFT} was constructed to capture the classical
dynamics described by the full Einstein-Maxwell theory for the scattering of
massive complex scalars interacting through the exchange of potential photons
and gravitons, the scattering amplitudes of both theories must be equal order
by order in the PM expansion
\begin{equation}
    \mathcal{M}^{\textrm{EFT}}_n  \stackrel{!}{=} \mathcal{M}_n \;.
\end{equation}
This matching procedure yields the following coefficient functions for the
potential
\begin{equation}
\begin{split}
    c_1(\boldsymbol{p}^2) &= \frac{\nu^2 m^2}{\gamma^2 \xi}(1-2\sigma^2+\eta_1\eta_2\sigma) \;, \\ 
    c_2(\boldsymbol{p}^2) &= \frac{\nu^2 m^3}{\gamma^2 \xi} \Bigg[\frac{3}{4}(1-5\sigma^2)-\frac{1}{2}\eta^2_1\eta^2_2+3\eta_1\eta_2\sigma-\frac{1}{4}(1-3\sigma^2)\Big(\frac{\eta^2_1+\eta^2_2}{2}+\frac{\eta^2_1-\eta^2_2}{2}\sqrt{1-4\nu}\Big)\Bigg] \\
    &+ \frac{\nu^3 m^3}{\gamma^3\xi^2}\Bigg[-4\sigma(1-2\sigma^2)+\eta^2_1\eta^2_2\sigma+\eta_1\eta_2(1-6\sigma^2)\Bigg] \\ 
    &+ \frac{\nu^4 m^3}{\gamma^5\xi^3}(1-\xi) \Bigg[-\frac{1}{2}(1-2\sigma^2)^2-\frac{1}{2}\eta^2_1\eta^2_2\sigma^2-\eta_1\eta_2\sigma(1-2\sigma^2)\Bigg] \;.
\end{split}
\label{eqn:coefficients}
\end{equation}

Let us note that the matching coefficients are, as expected, fully finite. The
cancellation of infrared divergent and superclassical contributions is made
apparent in the EFT amplitude. The superclassical and divergent piece at the
2PM order, the third term in Eq.~\eqref{eqn:EFT_2PM_amplitude}, is completely
determined by the lower order coefficient and therefore shows that this term is
indeed an iteration term and therefore does not contain any new classical
information about the two-body system.

\section{Classical Hamiltonian and Observables}
\label{sec:observables}
From the matching procedure we obtained the coefficient functions of the
effective two-body potential describing the classical conservative dynamics of
charged, compact binary systems. The 2PM center-of-mass Hamiltonian in
isotropic coordinates is given by
\begin{equation}
	H(\boldsymbol{p},\boldsymbol{r}) = \sqrt{\boldsymbol{p}^2+m^2_1} +  \sqrt{\boldsymbol{p}^2+m^2_2} + \sum_{n=1}^{2}c_n(\boldsymbol{p}^2)\left(\frac{G}{|\boldsymbol{r}|}\right)^n \;,
\label{eq:Hamiltonian}
\end{equation}
where $\boldsymbol{r}$ is the relative position-vector between the two compact
objects, which is the Fourier conjugate variable to the momentum transfer
$\boldsymbol{k}-\boldsymbol{k'} = \boldsymbol{q}$ in Eq. \eqref{eqn:potential},
and the coefficients $c_n(\boldsymbol{p}^2)$ are given in Eqs.
\eqref{eqn:coefficients}.  This Hamiltonian was determined by considering the
scattering of two electrically charged, massive spinless particles.
Nevertheless, it is valid not only for hyperbolic orbits but also for bound
orbits. We proceed to perform consistency checks against known results in the
literature.

\subsection{Point charge in a Reissner-Nordström background}
We start by checking that in the probe limit $m_2 \ll m_1$ our Hamiltonian
coincides with the Hamiltonian for a point charge in a Reissner-Nordström
background at 2PM order.

The motion of a massive charged relativistic particle of mass $\mu$ and charge
$eq$ in an electromagnetic field and in a spacetime with metric $g_{\mu\nu}$ is
determined by the Lagrangian (see e.g. Ref. \cite{1998mechanics})
\begin{equation}
    L(x^{\mu},\dot{x}^{\mu}) = -\mu\sqrt{g_{\mu\nu}\dot{x}^{\mu}\dot{x}^{\nu}} - eqA_{\mu}(x^{\mu})\dot{x}^{\mu} \;,
\end{equation}
where we defined $\dot{x}^\mu \equiv dx^\mu/d\tau$ with $\tau$ referring to the
proper time.  We are interested in the case where $g_{\mu\nu}$ refers to the
Reissner-Nordström metric of Eq. \eqref{eqn:Reissner-Nordstrom}. The metric
components $g_{i0}$ explicitly vanish. Moreover, we choose a fixed reference
frame with a worldline static gauge $x^{0} = \tau$. The Lagrangian then reads
\begin{equation}
    L(x^{\mu},\dot{x}^{\mu}) = -\mu\sqrt{g_{00}+g_{ij}\dot{x}^{i}\dot{x}^{j}} - eq A_0 + eqA_{i}\dot{x}^{i} \;.
\end{equation}
The Reissner-Nordström metric has a timelike Killing vector associated with
time translation invariance since the metric does not explicitly depend on
time. This implies the existence of a conserved quantity, namely the energy of
the particle along a geodesic $x^{\mu}(\tau)$. We can therefore set up the
Hamiltonian formulation of our problem. The conjugate momenta are 
\begin{equation}
    p_{i} \equiv \frac{\partial L}{\partial \dot{x}^{i}} = -\mu \frac{g_{ij}\dot{x}^{j}} {\sqrt{g_{00}+g_{ij}\dot{x}^{i}\dot{x}^{j}}} + eq A_i \;.
\end{equation}
The Hamiltonian is obtained in the usual fashion by taking the Legendre
transform of the Lagrangian. The result is
\begin{equation}
    H = p_i\dot{x}^{i} - L = \sqrt{g_{00}}\sqrt{\mu^2-g^{ij}p_ip_j} + eq A_0 \;.
\end{equation}
If we now substitute for the Reissner-Nordström metric of Eq.
(\ref{eqn:RN_isotropic}) and the electromagnetic four-potential of Eq.
(\ref{eqn:four-potential}), both in isotropic coordinates, we obtain what we
call the \textit{Reissner-Nordström Hamiltonian}
\begin{multline}
    H^{\textrm{RN}}(\boldsymbol{r},\boldsymbol{p}) = \mu \frac{\left(1-\frac{r_s}{4r}\right)\left(1+\frac{r_s}{4r}\right)+\frac{r^2_Q}{4r^2}}{\left(1+\frac{r_s}{4r}\right)^2-\frac{r^2_Q}{4r^2}}\sqrt{1+\left[\left(1+\frac{r_s}{4r}\right)^2-\frac{r_Q^2}{4r^2}\right]^{-2}\frac{\boldsymbol{p}^2}{\mu^2}} \\
+ \frac{e^2qQ}{4\pi r\left[\left(1+\frac{r_s}{4r}\right)^2-\frac{r_Q^2}{4r^2}\right]} \;.
\label{eqn:RN_Hamiltonian}
\end{multline}
In the neutral limit, this Hamiltonian reduces to the Hamiltonian for a test
particle in a Schwarzschild background as given in Ref. \cite{NWex_1993}. 

Let us now take the probe limit of this Hamiltonian. Our use of the symbols
$\mu$ (the reduced mass) to represent the point-charge mass and $m$ (the total
mass) for the mass of the Reissner-Nordström background was so that in the
probe limit, $m_2 \ll m_1$, we can make the replacements $\mu \to m_2$ and $m
\to m_1$. If we then perform a series expansion of Eq.
\eqref{eqn:RN_Hamiltonian} up to $\mathcal{O}(G^2)$, take the probe limit, and
perform a Fourier transform of the potential term we will obtain a potential in
the form of Eq. \eqref{eqn:potential} with the coefficients
\begin{equation}
\begin{split}
    c^{\textrm{RN}}_1(\boldsymbol{p}^2) &= -\frac{m_1}{E_2}(2\boldsymbol{p}^2+m^2_2) + m_1m_2\eta_1\eta_2 \;, \\
    c^{\textrm{RN}}_2(\boldsymbol{p}^2) &= \frac{m^2_1}{4E^3_2}(9\boldsymbol{p}^2+13m^2_2\boldsymbol{p}^2+2m^4_2) + \frac{m^2_1\eta^2_1}{4E_2}(3\boldsymbol{p}^2+2m^2_2) - m^2_1m_2\eta_1\eta_2 \;,
\end{split}
\end{equation}
where we identified $\eta_1 = eQ/(\sqrt{4\pi G}m_1)$ and $\eta_2 =
eq/(\sqrt{4\pi G}m_2)$.  These coefficients agree with those in Eq.
\eqref{eqn:coefficients} after taking the probe limit of the latter.

\subsection{Scattering Angle}
To obtain the conservative scattering angle for two charged black holes from
the Hamiltonian in Eq. \eqref{eq:Hamiltonian} we write down Hamilton's equations
of motion. Since the Hamiltonian is isotropic and does not contain terms
involving $\boldsymbol{p} \cdot \boldsymbol{r}$ but only $\boldsymbol{p}^2$,
Hamilton's equations imply
\begin{align}
    \dot{\boldsymbol{r}} &= \frac{\partial H(\boldsymbol{p}^2,\boldsymbol{r}^2)}{\partial \boldsymbol{p}} = \left[\frac{\partial H(\boldsymbol{p}^2,\boldsymbol{r}^2)}{\partial \boldsymbol{p}^2}\right]2\boldsymbol{p} \;, 
\label{eq:Hamilton's eq for coordinate}
    \\
    \dot{\boldsymbol{p}} &= -\frac{\partial H(\boldsymbol{p}^2,\boldsymbol{r}^2)}{\partial \boldsymbol{r}} = \left[-\frac{\partial H(\boldsymbol{p}^2,\boldsymbol{r}^2)}{\partial \boldsymbol{r}^2}\right]2\boldsymbol{r} \;.
\label{eq:Hamilton's eq for momentum}
\end{align}
The central-field potential $V(\boldsymbol{r}^2,\boldsymbol{p}^2)$ confines the
dynamics to a plane which we parametrize by polar coordinates $(r,\theta)$. In
these coordinates, vectors are decomposed into a sum of radial and tangential
parts; the relative position vector is $\boldsymbol{r} = r \boldsymbol{e}_r$
and the momentum is $\boldsymbol{p} = p_r \boldsymbol{e}_r +
p_{\theta}\boldsymbol{e}_{\theta}$, where $\boldsymbol{e}_r$ and
$\boldsymbol{e}_{\theta}$ are unit vectors in the radial and tangential
directions, respectively. 

Since the potential has spherical symmetry, the total energy $E$ and angular
momentum $J$ are conserved throughout the scattering process,
\begin{equation}
    E = H(r^2,p^2_r+p^2_{\theta})\;, \quad J = |\boldsymbol{r} \times \boldsymbol{p}| = rp_{\theta}\;.
\end{equation}
Since by assumption the potential $V$ vanishes as $r \to \infty$, the total
energy is the same as the kinetic energy of the system at infinity 
\begin{equation}
    E = \sqrt{p^2_{\infty}+m^2_1} + \sqrt{p^2_{\infty}+m^2_2} \;,
\label{eq:total energy}
\end{equation}
and the angular momentum at infinity is given by
\begin{equation}
    J = bp_{\infty} \;,
\end{equation}
where $b = |\boldsymbol{b}|$ is the magnitude of the impact parameter. Equation
\eqref{eq:Hamilton's eq for coordinate} states that $\dot{\boldsymbol{r}}$ and
$\boldsymbol{p}$ are parallel. Using the chain rule and the conservation of
angular momentum, this determines the scattering trajectory in terms of the
radial momentum
\begin{equation}
    \frac{\mathrm{d}\theta}{\mathrm{d}r} = \frac{\mathrm{d}\theta}{\mathrm{d}t}\frac{\mathrm{d}t}{\mathrm{d}r} = \frac{p_{\theta}}{rp_r} = \frac{J}{r^2p_r}\;.
\end{equation}
The scattering angle is then obtained by integrating the above equation:
\begin{equation}
    \chi = -\pi + 2J \int_{r_\textrm{min}}^{\infty} \frac{\mathrm{d}r}{r^2 p_r(r)} \;,
\label{eq:scattering angle}
\end{equation}
where $p_r$ as a function of $r$ is determined by the conservation of energy $E
= H(r^2,p^2_r+J^2/r^2)$, and $r_{\textrm{min}}$ denotes the minimum distance
between the two particles and is the point at which the radial momentum
vanishes,
\begin{equation}
    p_r(r_{\textrm{min}}) = 0\;.
\label{eq:minimum distance definition}
\end{equation}
From the conservation of energy and Eq. \eqref{eq:total energy} it follows that
\begin{multline}
    \sqrt{p^2_{\infty}+m^2_1} + \sqrt{p^2_{\infty}+m^2_2} \\= \sqrt{p^2_r+J^2/r^2+m^2_1} +     \sqrt{p^2_r+J^2/r^2+m^2_2} + \sum_{n=1}^{2}c_n(p^2_r+J^2/r^2)\left(\frac{G}{r}\right)^n \;,
\end{multline}
which we can solve for the radial momentum as a function of $r$ perturbatively
in $G$. To order $G^2$, we obtain
\begin{equation}
    p^2_r(r) = p^2_{\infty} - \frac{J^2}{r^2} + \sum_{n=1}^{2} P_n \left(\frac{G}{r}\right)^n \;,
\label{eq:radial momentum}
\end{equation}
with the coefficients in the series given by 
\begin{align}
    P_1 &= -2E\xi\bar{c}_1 \;,  \label{eq:P1}\\
    P_2 &= -2E\xi\bar{c}_2 + (1-3\xi)\bar{c}^2_1 + 4E^2\xi^2\bar{c}_1\bar{c}'_1\;,
\label{eq:P2}
\end{align}
where $\bar{c}_n = c_n(p^2_{\infty})$ and the prime denotes a derivative with
respect to the argument. The minimum distance is then obtained by solving Eq.
\eqref{eq:minimum distance definition} perturbatively in $G$, which to order
$G^2$ gives
\begin{equation}
    r^2_{\textrm{min}} = b^2 - b^3P_1\frac{G}{J^2} + \frac{b^4}{2}(P^2_1-2p^2_{\infty}P_2)\frac{G^2}{J^4} + \mathcal{O}(G^3)\;.
\label{eq:minimum distance}
\end{equation}
Using Eqs. \eqref{eq:radial momentum} and \eqref{eq:minimum distance} in Eq.
\eqref{eq:scattering angle}, the scattering angle through second order in the
PM expansion is given by \cite{Bern:2019crd}
\begin{equation}
    \chi = \sum_{i \geq1} \chi^{i\textrm{PM}} = \frac{P_1}{p_{\infty}}\left(\frac{G}{J}\right) + \frac{\pi}{2}P_2\left(\frac{G}{J}\right)^2 + \mathcal{O}\left((G/J)^3\right)\;.
\label{eq:PM expansion of scattering angle}
\end{equation}
Using Eqs. \eqref{eq:P1} and \eqref{eq:P2} along with the coefficient functions
in \eqref{eqn:coefficients}, the conservative scattering angle up to
$\mathcal{O}(G^2)$ is
\begin{equation}
\begin{split}
    \chi^{1\textrm{PM}} &= \frac{2(2\sigma^2-1-\eta_1\eta_2\sigma)}{\sqrt{\sigma^2-1}}\frac{Gm^2\nu}{J} \;, \\
    \chi^{2\textrm{PM}} &= \frac{\pi}{4\gamma}\left[3(5\sigma^2-1)+2\eta^2_1\eta^2_2-12\eta_1\eta_2\sigma+\left(\frac{\eta^2_1+\eta^2_2}{2}+\frac{\eta^2_1-\eta^2_2}{2}\sqrt{1-4\nu}\right)(1-3\sigma^2)\right]\frac{G^2m^4\nu^2}{J^2} \;.
\end{split}
\label{eq:scattering angle at 1PM and 2PM}
\end{equation}
We find perfect agreement between our result and the 2PM scattering angle
provided in Ref.~\cite{Wilson-Gerow:2023syq} which was obtained using an
effective theory for extreme mass ratios. In the neutral limit of vanishing
electric charges, we correctly reproduce the 2PM scattering angle in Ref.
\cite{Bern:2019crd}, and in the limit $G \to 0$ the 2PL scattering angle in
Ref. \cite{Bern:2021xze}. 

\subsection{Post-Newtonian expansion and observables for bound systems}
In this section we compute the post-Newtonian expansion of the scattering angle
as well as the binding energy and periastron advance --- two adiabatic
invariants --- for a charged black hole binary in \textit{bound} orbits.
Instead of calculating these observables using the Hamiltonian in Eq.
(\ref{eq:Hamiltonian}), we follow Refs. \cite{Kalin:2019rwq,Kalin:2019inp}
where an explicit dictionary is introduced relating scattering data and
observables for bound states.

The dictionary is based on a relation between the relative momentum and the
scattering amplitude that is presumed to hold to all orders in the PM
expansion, namely\footnote{We are ignoring radiation-reaction effects, which
include tail effects, since they start appearing at 4PM order.}
\begin{equation}
    \boldsymbol{p}^2 = p^2_{\infty}+ \frac{1}{2E}\int \mathrm{d}^3\boldsymbol{r}\mathcal{M}(\boldsymbol{p},\boldsymbol{q})e^{i\boldsymbol{q}\cdot\boldsymbol{r}}\;,
\end{equation}
where $\mathcal{M}(\boldsymbol{p},\boldsymbol{q})$ represents the IR-finite
part of the relativistically normalized scattering amplitude in the classical
limit. The above equation is then expanded perturbatively as 
\begin{equation}
    \boldsymbol{p}^2 = p^2_{\infty}+\sum_{n=1}\widetilde{\mathcal{M}}_n\frac{G^n}{r^n}\;,
\end{equation}
which when compared with Eq. (\ref{eq:radial momentum}), extended to all PM
orders, gives the relation
\begin{equation}
    \widetilde{\mathcal{M}}_n = P_n \equiv p^2_{\infty}m^nf_n\;.
\end{equation}
The coefficients $f_n$ in the expansion of the amplitude are the ones that
appear more naturally in the PM expansion of the scattering angle (\ref{eq:PM
expansion of scattering angle})
\begin{equation}
\begin{split}
    \chi &= p_{\infty}f_1\frac{Gm}{J} + \frac{\pi}{2}p^2_{\infty}f_2\left(\frac{Gm}{J}\right)^2+p^3_{\infty}\left(2f_3+f_2f_1-\frac{f^3_1}{12}\right)\left(\frac{Gm}{J}\right)^3 
    \\
    & \quad + \frac{3\pi}{8}p^4_{\infty}(f^2_2+2f_1f_3+2f_4)\left(\frac{Gm}{J}\right)^4 + \cdots \;,
\label{eq:Appendix scattering angle}
\end{split}
\end{equation}
where we have explicitly written the expansion of the scattering angle to order
$G^4$ for later use. If we now compare with our result in Eq.
(\ref{eq:scattering angle at 1PM and 2PM}) and use the relation
\begin{equation}
    p_{\infty} = \frac{m\nu}{\gamma}\sqrt{\sigma^2-1}\;,
\label{eq:momentum at infinity}
\end{equation}
we can extract the coefficients $f_1$ and $f_2$ for Einstein-Maxwell theory,
\begin{equation}
\begin{split}
    f_1 &= 2\gamma \left(\frac{2\sigma^2-1-\eta_1\eta_2\sigma}{\sigma^2-1}\right) \;, \\
    f_2 &= \frac{\gamma}{2(\sigma^2-1)}\left[3(5\sigma^2-1)+2\eta^2_1\eta^2_2-12\eta_1\eta_2\sigma+\left(\frac{\eta^2_1+\eta^2_2}{2}+\frac{\eta^2_1-\eta^2_2}{2}\sqrt{1-4\nu}\right)(1-3\sigma^2)\right] \;.
\end{split}
\end{equation}
In order to obtain results at 2PN accuracy we need to go to 3PM, \textit{i.e.},
include the $f_3$ contribution. From the result of the scattering angle at 3PM
\cite{Wilson-Gerow:2023syq}, we obtain that
\begin{equation}
\begin{split}
    f_3 &= \frac{\gamma}{6(\sigma^2-1)^4}\Bigg\{2\gamma^2(\sigma^2-1)(2\sigma^2-1-\eta_1\eta_2\sigma)^3-(1-2\nu)(\sigma^2-1)\Bigg[10-3(\eta^2_1+\eta^2_2)-18\eta^2_1\eta^2_2 
    \\
    &-3\eta_1\eta_2\sigma(-30+3\eta^2_1+3\eta^2_2+2\eta^2_1\eta^2_2)+3\sigma^2(-40+9\eta^2_1+9\eta^2_2+24\eta^2_1\eta^2_2) 
    \\
    &+\eta_1\eta_2\sigma^3(-240+21\eta^2_1+21\eta^2_2+4\eta^2_1\eta^2_2) -48\sigma^4(-5+\eta^2_1+\eta^2_2+\eta^2_1\eta^2_2)-12\eta_1\eta_2\sigma^5(-12+\eta^2_1+\eta^2_2)
    \\
    &+8\sigma^6(-16+3\eta^2_1+3\eta^2_2)\Bigg] -3\gamma(\sigma^2-1)^2(2\sigma^2-1-\eta_1\eta_2\sigma)\Bigg[3(5\sigma^2-1)+2\eta^2_1\eta^2_2-12\eta_1\eta_2\sigma 
    \\
    &+\left(\frac{\eta^2_1+\eta^2_2}{2}+\frac{\eta^2_1-\eta^2_2}{2}\sqrt{1-4\nu}\right)(1-3\sigma^2)\Bigg]-3\sqrt{1-4\nu}(\sigma^2-1)^2\Bigg[(\eta^2_1-\eta^2_2)(1-8\sigma^2+8\sigma^4)
    \\
    &+\eta_1\eta_2\sigma(\eta^2_1-\eta^2_2)(3-4\sigma^2)\Bigg] 
    + 2\nu(\sigma^2-1)\Bigg[2\eta^3_1\eta^3_2(3-3\sigma^2+\sigma^4)+6\eta^2_1\eta^2_2\sigma(-1-4\sigma^2+4\sigma^4) 
    \\
    &+ \eta_1\eta_2(16-42\sigma^2+96\sigma^4-64\sigma^6)+\eta_1\eta_2(\eta^2_1+\eta^2_2)(7-9\sigma^2-6\sigma^4+8\sigma^6)
    \\
    &+(\eta^2_1+\eta^2_2)(23\sigma-51\sigma^3+36\sigma^5-8\sigma^7)+2\sigma(-55+132\sigma^2-114\sigma^4+36\sigma^6)\Bigg] +6\nu(\sigma^2-1)^{5/2} \Bigg[12
    \\
    &+48\sigma^2-16\sigma^4+16\eta_1\eta_2\sigma(\sigma^2-2)-4(\eta^2_1+\eta^2_2)(2\sigma^2+1)+4\eta^2_1\eta^2_2(1-2\sigma^2)\Bigg] \mathrm{arccosh}(\sigma)
    \Bigg\} \;.
\end{split}
\end{equation}
As expected, there is perfect agreement between Eqs. (5.60) and (5.65) of Ref.
\cite{Kalin:2019rwq} for Einstein gravity and the $f_n$ coefficients above upon
taking the neutral limit. 

In Ref. \cite{Kalin:2019inp}, a surprisingly simple relation between the
scattering angle and the periastron advance was established, valid to all
orders in the PM expansion. The relation is given by
\begin{equation}
    \Delta\Phi = \chi(J) + \chi(-J) \;,
\end{equation}
and substituting Eq. (\ref{eq:Appendix scattering angle}) we find that
\begin{equation}
    \Delta\Phi = \pi\hat{p}^2_{\infty}f_2\left(\frac{1}{j^2}\right)+\frac{3\pi}{4}\hat{p}^4_{\infty}(f^2_2+2f_1f_3+2f_4)\left(\frac{1}{j^4}\right) + \cdots \;,
\label{eq:periastron advance}
\end{equation}
where $\hat{p}_{\infty} = p_{\infty}/(m\nu)$ and we introduced the rescaled
angular momentum $j=J/(Gm^2\nu)$. Defining the binding energy as
\begin{equation}
    \mathcal{E} = \frac{E-m}{m\nu} \;,
\end{equation}
in terms of which it is easy to show that 
\begin{equation}
\begin{split}
    \sigma &\equiv \frac{E^2-m^2_1-m^2_2}{2m_1m_2} = 1+\mathcal{E}+\frac{1}{2}\nu\mathcal{E}^2 \;, \\
    \gamma &\equiv E/m = \sqrt{1+2\nu(\sigma-1)} = 1+\nu\mathcal{E}\;.
\label{eq:sigma and gamma}
\end{split}
\end{equation}
Using Eq. (\ref{eq:sigma and gamma}) to write the $f_n$'s as a function of the
binding energy, substituting the expressions thereof in Eq. (\ref{eq:periastron
advance}), and performing a Taylor series expansion in powers of the binding
energy, we obtain
\begin{equation}
\begin{split}
   K &\equiv 1+\frac{\Delta\Phi}{2\pi} = 1+\frac{1}{j^2}\Bigg[3-\frac{\eta^2_1}{4}(1+X_{12})-3\eta_1\eta_2-\frac{\eta^2_2}{4}(1-X_{12})+\frac{\eta^2_1\eta^2_2}{2}\Bigg]\frac{1}{c^2} \\
    &+\Bigg\{\Bigg[\frac{15}{2}-3\nu+\frac{\eta^2_1}{4}(1+X_{12})(-3+\nu)+3\eta_1\eta_2(-1+\nu)+\frac{\eta^2_2}{4}(1-X_{12})(-3+\nu)-\frac{\nu}{2}\eta^2_1\eta^2_2\Bigg]\frac{\mathcal{E}}{j^2} \\
    &+\Bigg[\frac{105}{4}-\frac{15}{2}\nu+\frac{15}{2}\eta_1\eta_2(-7+3\nu)+\frac{3}{4}\eta^2_1\eta^2_2(45-36\nu)+\frac{15}{2}\eta^3_1\eta^3_2(-1+2\nu)+\frac{3}{8}\eta^4_1\eta^4_2(1-6\nu) \\
    &+\frac{3\eta^4_1}{32}(1+X_{12})^2-\frac{45}{8}\eta^2_1(1+X_{12})-\frac{3\nu}{8}\eta^2_1(1-X_{12})+\frac{15}{2}\eta^3_1\eta_2(1+X_{12})+\frac{3\nu}{4}\eta^3_1\eta_2(3-X_{12}) \\
    &+\frac{3}{8}\eta^4_1\eta^2_2X_{12}(-6+\nu)-\frac{3}{8}\eta^4_1\eta^2_2(6+5\nu) + \frac{3\eta^4_2}{32}(1-X_{12})^2-\frac{45}{8}\eta^2_2(1-X_{12}) -\frac{3\nu}{8}\eta^2_2(1+X_{12}) \\
    &+\frac{15}{2}\eta_1\eta^3_2(1-X_{12})+\frac{3\nu}{4}\eta_1\eta^3_2(3+X_{12})-\frac{3}{8}\eta^2_1\eta^4_2X_{12}(-6+\nu)-\frac{3}{8}\eta^2_1\eta^4_2(6+5\nu) 
    \Bigg]\frac{1}{j^4}\Bigg\}\frac{1}{c^4} \;,
\end{split}
\label{eq:Periastron}
\end{equation}
where we define the ($m$-rescaled) mass difference $X_{12} = \sqrt{1-4\nu} =
(m_1-m_2)/m$, and we have restored factors of $c$ to indicate the PN order by
the formal PN-expansion parameter $1/c^2$. The complete expression is symmetric
under the interchange of charges $\eta_1 \leftrightarrow \eta_2$ and the
replacement $X_{12}\rightarrow -X_{12}$.
Equation (\ref{eq:Periastron}) is the 2PN-accurate expression for the periastron
advance and in the neutral limit reproduces the known 2PN result in general
relativity, see \textit{e.g.} Eq. (C26) in Ref. \cite{Bini:2017wfr}. Of course,
the full expression in Eq. (\ref{eq:periastron advance}) contains additional PN
corrections beyond 2PN order. In particular, the $\mathcal{O}(1/j^2)$
contribution is exact to all orders in the binding energy, though we would
still need the $f_4$ coefficient (three-loop amplitude) to give the complete
$1/j^4$ term valid to all orders in binding energy. 

For the case of circular orbits, the following relation between the rescaled
angular momentum and the scattering data holds at 3PM order
\cite{Kalin:2019rwq}
\begin{equation}
    j^2 = \frac{1-\sigma^2}{1+2\nu(\sigma-1)}\left[\left(\frac{f_1}{2}\right)^2-f_2+\frac{2f_3}{f_1}\right] + \cdots \;,
\label{eq:3PM reduced angular momentum}
\end{equation}
where we have truncated the full expression to only the terms we need for 2PN
accuracy. Substituting the expressions for the $f_n$'s written as functions of
the binding energy one can then compute the orbital frequency from the first
law of binary black hole mechanics \cite{LeTiec:2011ab}
\begin{equation}
    Gm\Omega = \left(\frac{\mathrm{d}j(\mathcal{E})}{\mathrm{d}\mathcal{E}}\right)^{-1} \;.
\label{eq:first law BBH mechanics}
\end{equation}
If we now introduce the so-called charge-flexed frequency parameter of Ref.
\cite{Placidi:2025xyi}
\begin{equation}
    x_q \equiv \left[\frac{Gm\Omega}{c^3}(1-\eta_1\eta_2)\right]^{2/3} \;,
\end{equation}
we see that Eq. (\ref{eq:first law BBH mechanics}) allows us to express this
parameter as a function of the binding energy, $x_q(\mathcal{E})$. By
performing a Taylor series expansion in powers of $\mathcal{E}$ and inverting
the series, we then obtain the following expression for the binding energy at
2PN order
\begin{equation}
\begin{split}
    \mathcal{E}(x_q) &= -\frac{x_q}{2} \\
    & +\frac{x^2_q}{24}(1-\eta_1\eta_2)^{-2} \Bigg\{9+\nu -2(1+X_{12})\eta^2_1 -2(3+\nu)\eta_1\eta_2  -2(1-X_{12})\eta^2_2+(1+\nu)\eta^2_1\eta^2_2 \Bigg\}\frac{1}{c^2} 
    \\
    &+  \frac{x^3_q}{48}(1-\eta_1\eta_2)^{-4} \Bigg\{ 81 -57\nu +\nu^2  -18(1+X_{12})\eta^2_1 -2\nu(5-X_{12})\eta^2_1  - 2(81-84\nu+2\nu^2)\eta_1\eta_2 \\
   & -18(1-X_{12})\eta^2_2 -2\nu(5+X_{12})\eta^2_2 -2(1+X_{12})^2\eta^4_1 +4\eta^3_1\eta_2\Big[9  +11\nu +(9 -\nu)X_{12}\Big] \\
   &  +(90-254\nu+6\nu^2)\eta^2_1\eta^2_2 +4\eta_1\eta^3_2\Big[9+11\nu -(9 -\nu)X_{12}\Big] -2(1-X_{12})^2\eta^4_2 \\
   &  +2\eta^2_1\eta^4_2(2\nu-19-9X_{12}) +2\eta^2_1\eta^4_2\Big[14(1+X_{12})-\nu(19+X_{12})\Big] +2\eta^4_1\eta^2_2(2\nu-19+9X_{12})   \\
   &+2\eta^4_1\eta^2_2\Big[14(1-X_{12})-\nu(19-X_{12})\Big] -  2 \eta^3_1\eta^3_2(9-76\nu+2\nu^2)+(1-25\nu+\nu^2)\eta^4_1\eta^4_2 \Bigg\}\frac{1}{c^4} \;.
\end{split}
\label{eq:2PN binding energy}
\end{equation}
Next, by taking Eq. (\ref{eq:3PM reduced angular momentum}) and plugging it in
Eq. (\ref{eq:Periastron}) we can obtain the periastron advance as a function of
the binding energy. If we then substitute Eq. (\ref{eq:2PN binding energy}) we
obtain the periastron advance as a function of the $x_q$. To 2PN order, we
obtain 
\begin{equation}
\begin{split}
    K &=  1 + \frac{x_q}{4}(1-\eta_1\eta_2)^{-2}\Bigg[12-\eta^2_1-12\eta_1\eta_2-\eta^2_2+2\eta^2_1\eta^2_2+X_{12}(\eta^2_2-\eta^2_1)\Bigg]\frac{1}{c^2}
    \\
    & +\frac{x^2_q}{48}(1-\eta_1\eta_2)^{-4}\Bigg\{648 -120(1+X_{12})\eta^2_1-\frac{7}{2}(1+X_{12})^2\eta^4_1 -1368\eta_1\eta_2 + 204(1+X_{12})\eta^3_1\eta_2
    \\
    & -120(1-X_{12})\eta^2_2 +4(222-7\nu)\eta^2_1\eta^2_2 -70(1+X_{12})\eta^4_1\eta^2_2 + 204(1-X_{12})\eta_1\eta^3_2 -192\eta^3_1\eta^3_2 
    \\
    & -\frac{7}{2}(1-X_{12})^2\eta^4_2 -70(1-X_{12})\eta^2_1\eta^4_2 +10\eta^4_1\eta^4_2 + \nu\Bigg[-336 -20\eta^2_1 +1008\eta_1\eta_2+112\eta^3_1\eta_2 -20\eta^2_2 
    \\
    & -1256\eta^2_1\eta^2_2 -92\eta^4_1\eta^2_2 +112\eta_1\eta^3_2 + 688\eta^3_1\eta^3_2  -92\eta^2_1\eta^4_2 -104\eta^4_1\eta^4_2 +16X_{12}\Bigg(\eta^2_1 -2\eta^3_1\eta_2 -\eta^2_2 + \eta^4_1\eta^2_2
    \\
    &+ 2\eta_1\eta^3_2 -\eta^2_1\eta^4_2
    \Bigg)\Bigg] \Bigg\}\frac{1}{c^4} \;.
\end{split}
\label{eq:2PN periastron advance}
\end{equation}
Lastly, we give the PN expansion of the 3PM scattering angle. We take Eq.
(\ref{eq:Appendix scattering angle}) truncated to 3PM order and substitute in
the expressions for $f_n$'s. We also use Eqs. (\ref{eq:momentum at infinity})
and (\ref{eq:sigma and gamma}) so that the scattering angle is written as a
function of the relativistic Lorentz factor, $\chi(\sigma)$. If we introduce
the reduced angular momentum $p_{\phi} = J/(m\nu)$ and the variable $v_{\infty}
= \sqrt{\sigma^2-1}$, which measures the relative velocity at infinite
separation (in units of $c$), we can Taylor expand the scattering angle in
powers of $v_{\infty}$. The result up to 2PN order is 
\newpage
\begin{equation}
    \begin{split}
        \chi &=  \frac{2Gm}{p_{\phi}v_{\infty}}(1-\eta_1\eta_2) - \frac{2G^3m^3}{3p^3_{\phi}v^3_{\infty}}(1-\eta_1\eta_2)^3 \\
        &  +\frac{1}{c^2}\Bigg\{\frac{Gmv_{\infty}}{p_{\phi}}(4-\eta_1\eta_2)+\frac{G^2m^2\pi}{p^2_{\phi}}\Bigg[3-\frac{\eta^2_1}{4}(1+X_{12})-3\eta_1\eta_2-\frac{\eta^2_2}{4}(1-X_{12})+\frac{\eta^2_1\eta^2_2}{2}\Bigg] 
        \\
        & + \frac{G^3 m^3}{p^3_{\phi}v_{\infty}}\Bigg[8 {\;-\eta^2_1(1+X_{12})}-15\eta_1\eta_2{\;+\eta^3_1\eta_2(1+X_{12})}-\eta^2_2(1-X_{12})+8\eta^2_1\eta^2_2{\;+\eta_1\eta^3_2(1-X_{12})}-\eta^3_1\eta^3_2\Bigg]\Bigg\}
        \\
        & +\frac{1}{c^4}\Bigg\{\frac{G^2m^2\pi v^2_{\infty}}{p^2_{\phi}}\Bigg[\frac{3}{4}(5-2\nu)+\frac{\eta^2_1}{8}(1+X_{12})(-3+\nu){\;+\frac{3}{2}(-1+\nu)\eta_1\eta_2+\frac{\eta^2_2}{8}(1-X_{12})(-3+\nu)}-\nu\frac{\eta^2_1\eta^2_2}{4}\Bigg]\\
        &  +\frac{G^3m^3v_{\infty}}{p^3_{\phi}}\Bigg[-16(-3+\nu)-\frac{17}{2}\eta^2_1(1+X_{12}){\;+\frac{\eta^2_1}{2}(1-2\nu+X_{12})}+\nu\eta^2_1(1+X_{12})+\frac{\eta_1\eta_2}{4}(128\nu-225) \\
        & +\eta^3_1\eta_2(1+X_{12})(6-\nu)-\frac{3}{2}\eta^3_1\eta_2(1-2\nu+X_{12})-\frac{17}{2}\eta^2_2(1-X_{12})+\frac{\eta^2_2}{2}(1-2\nu-X_{12})+\nu\eta^2_2(1-X_{12}) \\
        &  +(16-24\nu)\eta^2_1\eta^2_2 +\eta_1\eta^3_2(1-X_{12})(6-\nu)-\frac{3}{2}\eta_1\eta^3_2(1-2\nu-X_{12})+\frac{\eta^3_1\eta^3_2}{4}(-3+16\nu)\Bigg]
        \\
        &  
         +\frac{Gmv^3_{\infty}}{4p_{\phi}}\eta_1\eta_2\Bigg\} \;.
    \end{split}
\label{eq:2PN scattering angle}
\end{equation}

We have compared our results for the binding energy (\ref{eq:2PN binding
energy}), the periastron shift (\ref{eq:2PN periastron advance}), and the
scattering angle (\ref{eq:2PN scattering angle}) at 2PN order with the results
obtained in Ref. \cite{Placidi:2025xyi} and found agreement in each case. 

\section{Conclusions}
\label{sec:conclusions}
In this paper we have computed the conservative dynamics for classical
scattering of two spinless charged black holes in Einstein-Maxwell theory at
$\mathcal{O}(G^2)$. Specifically, using an EFT matching procedure, we extracted
the 2PM-accurate conservative Hamiltonian that governs the orbital dynamics of
charged compact binaries. We considered binary systems of arbitrary masses and
charges so we can interpolate between Einstein gravity and electromagnetism by
taking the appropriate limits. In both of these limits our Hamiltonian
reproduces known results \cite{Bern:2019crd, Bern:2021xze}.

In the probe limit, we have shown the equivalence of our Hamiltonian with the
Hamiltonian for a point charge in a Reissner-Nordström background. This
provides a check in the probe limit valid to all orders in velocity.
Furthermore, by computing the scattering angle we found perfect agreement with
the result previously reported in Ref. \cite{Wilson-Gerow:2023syq}. We also
found agreement with observables for bound systems at 2PN with Ref.
\cite{Placidi:2025xyi}.

The results we have reported in this paper are limited to the first two orders
in the PM expansion and obtaining higher orders is a natural continuation. Some
additional avenues for future work are also immediately apparent, \textit{e.g.}
to consider the scattering of higher-spin fields in order to incorporate the
effect of the initial angular momentum of the binary constituents on the
dynamics, as well as taking into account finite-size effects. Another future
direction involves incorporating the effects of radiation and computing
radiative classical observables such as the impulse, radiated momentum, and
radiated energy.

\subsection*{Acknowledgments}
We thank Dogan Akpinar for comments on the manuscript and Fernando Febres
Cordero for supporting us in the use of the \textsc{Caravel} framework.
We thank the authors of Ref.~\cite{Placidi:2025xyi} for providing a cross check
of our 2PN observables.
MK thanks Aaron Zimmerman for helpful discussions during the workshop
\textit{Gravitational Waves meet Amplitudes in the Southern Hemisphere} at ICTP
S\~{a}o Paulo.  
Financial support through the DGAPA-PAPIIT grants IA102224 and IA103126 at
UNAM, and the SECIHTI “Ciencia Básica y de Frontera” grant CBF-2025-I-615 is
acknowledged. A.A.A gratefully acknowledges the financial support from the
Secretaría de Ciencia, Humanidades, Tecnología e Innovación. 

\bibliography{ref}
\end{document}